# General laws of reflection and refraction for subwavelength phase grating

Zhaona Wang [1]*, Yanyan Sun [1], Lu Han [1], Dahe Liu [1,2] *

1 Applied Optics Beijing Area Major Laboratory, Department of Physics,
Beijing Normal University, Beijing 100875

2 Key Laboratory of Nondestructive Test (Ministry of Education), Nanchang Hangkong University,
Nanchang 330063, China.

* Corresponding authors: zhnwang@bnu.edu.cn for Zhaona Wang and dhliu@bnu.edu.cn for Dahe Liu

**ABSTRACT**

The general reflection and refraction laws at the metasurface with the abrupt phase shift were derived by two different methods of Fermat's principle and the boundary conditions respectively. It is found that one or two critical angles for total internal reflection exist when a light hits on the optical sparse material from the optical denser material, and one such a critical angle exists when light spreads from the optical sparse material to the optical denser material. Anomalous reflection and refraction, such as, negative reflection and negative refraction may occur when a light passes through the metasurface, and the conditions of their occurrence were given. Finally, a kind of metasurface based on one-dimensional phase mask was designed to control the light propagation.

*OCIS Codes*: Metasurface; refraction law; reflection law; phase discontinuity

## 1. Introduction

The laws of reflection and refraction are the basis of geometrical optics. The behavior of a light beam at the interface is completely determined by the optical properties of the two materials at different side of the interface. In classical optics, it is assumed that the physical interface has an ideal boundary which does not change any information of light waves.

With the recent development of nanofabrication technology, metamaterials [1-3] were used to control the electromagnetic field, with which some properties that may not be found in nature can be obtained, such as, negative refraction index [4-6], ultra-magnetic property [7], low-loss property [8] and so on. Metasurface, as an important branch of metamaterials, has been used to achieve arbitrary control of light waves due to its special property, for example, to control the polarization state [9], to achieve completely transmission [10], to improve the antenna radiation performance [11-13], to obtain super absorption [14].

Different from the ideal classical boundary, metasurface can change the wavefront, the phase and the polarization state of a light wave, and the propagation of a light wave at the interface may violate the conventional reflection and refraction laws [15,16]. In order to further understand the behavior of light at the metasurface and to further design the devices with metasurface, it is necessary to derive the general laws of reflection and refraction at the metasurface which can generate the abrupt phase shift for a light wave. In the present work, the general laws of reflection and refraction were derived by the two methods based on Fermat's principle and the continuous boundary conditions, respectively. The relevant factors influencing light propagation were discussed. The full transmission critical condition, the total reflection critical condition, the conditions of abnormal reflection and abnormal refraction were also demonstrated. Finally, a metasurface made by a one-dimensional phase mask was designed to control the propagation of light.

## 2. Derivation from Fermat's principle

Fermat's principle is the basic principle of optics. It should be obeyed in transformation optics [17,18] although one needs to explore unusual geometries of space-time in the early universe [19]. Fermat's principle states that the ray trajectory of light between two points A and B is that of the extreme value of optical path, i.e. $\delta \int_A^B n(\vec{r}) dr = 0$, where $n(\vec{r})$ is the local refractive index. For a light wave, changing the optical path means actually changing the phase of the light wave. Therefore, Fermat's principle can also be stated as $\delta \int_A^B d\varphi(\vec{r}) = 0$, the principle of stationary phase [20–22], that is to say, the total phase $\int_A^B d\varphi(\vec{r})$ accumulated along the actual light path from point A to point B will be a constant. If an abrupt phase shift $\varphi(\vec{r}_s)$ depending on the coordinate $\vec{r}_s$ along the interface over the scale of the wavelength is introduced, then, the total phase shift $\Phi(\vec{r}_s) + \int_A^B \vec{k} \cdot d\vec{r}$ will be stationary, $\Phi(\vec{r}_s)$ respects the phase mutation caused by the metasurface at the point $\vec{r}_s$, $\int_A^B \vec{k} \cdot d\vec{r}$ is the phase change of light wave accumulated by gradual phase changes along the optical path, $\vec{k}$ is the wave vector of the propagating light.

Assume that the phase shift gradient along the interface $d\Phi/dx$ is a nonzero constant, the general reflection law derived from Fermat's principle had been given in [15]:

$$n_t \sin(\theta_t) - n_i \sin(\theta_i) = \frac{\lambda_0}{2\pi} \frac{d\Phi}{dx} \quad . \tag{1}$$

where $\theta_i$ is the incident angle, $\theta_t$ is the refractive angle; $n_i$ and $n_t$ are the refractive indices of the incident medium and the output medium, respectively. $\lambda_0$ is the wavelength of light in vacuum. From Eq. (1), the refracted beam can be in arbitrary direction by introducing a suitable phase discontinuity $d\Phi/dx$ along the interface. Because of the nonzero phase shift gradient, the two incident angles $\pm\theta_i$ will lead to different refractive angle. That is to say, the symmetry for the incident angle will be destroyed. The critical angle for total internal reflection will be:

$$\theta_{tc} = \arcsin(\pm\frac{n_t}{n_i} - \frac{\lambda_0}{2\pi n_i}\frac{d\Phi}{dx}) \quad (2)$$

From formula (2), it can be found that the critical angle for total internal reflection is not only related to the refractive indices of the two materials and the wavelength of light in vacuum, but also to the phase shift gradient at the interface. It means that, when the light wave is incident on the metasurface from the optical denser material to the optical sparse material (i.e. $n_t < n_i$), there will be two critical angles for total internal reflection with the condition of $-\frac{n_t}{n_i} - \frac{\lambda_0}{2\pi n_i}\frac{d\Phi}{dx} \geq -1$, (i.e. $\frac{d\Phi}{dx} \leq (-n_t + n_i)\frac{2\pi}{\lambda_0}$) and one critical angles for total reflection with the condition of $\frac{d\Phi}{dx} > (-n_t + n_i)\frac{2\pi}{\lambda_0}$. It is different from the prediction in Ref. [15] in which there always exist two critical angles. Meanwhile, quite different from the classical refraction law, the general refraction law also allows the critical angle for total internal reflection when light beam is incident from the optical sparse material to the optical denser material (i.e. $n_t > n_i$) with the conditions of $\frac{d\Phi}{dx} \geq (n_t - n_i)\frac{2\pi}{\lambda_0}$.

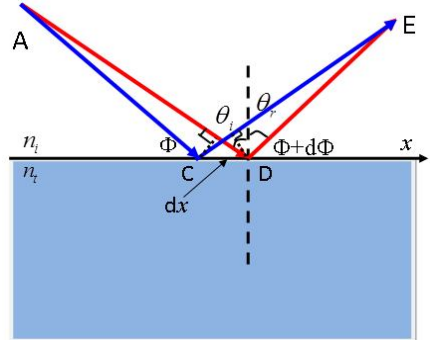

Fig. 1. Schematics used to derive the generalized Snell's law of reflection. The interface between the two media is artificially structured in order to introduce an abrupt phase shift in the light path, which is a function of the position along the interface. $\Phi$ and $\Phi + d\Phi$ are the phase shifts where the two paths (blue and red) cross the boundary.

Now, we derived the general reflection law. Considering a plane wave with a incident angle of $\theta_i$, assuming that the two paths ACE and ADE are infinitesimally close to the actual path between the points A and E (See Fig. 1), then the phase difference between them should be zero.

$$[k_0 n_i \sin(\theta_i)dx + (\Phi + d\Phi)] - [k_0 n_i \sin(\theta_r)dx + \Phi] = 0 \quad (3)$$

where $\theta_r$ is the reflected angle, respectively, $\Phi$ and $\Phi + d\Phi$ are the phase discontinuities at the locations where the two paths just cross the interface, $dx$ is the distance between the two cross points, and the wave number $k_0 = 2\pi/\lambda_0$. Assuming the phase shift gradient along the interface $d\Phi/dx$ is chosen to be a nonzero constant, the Eq. (3) will leads to the general reflection law

$$\sin(\theta_r) - \sin(\theta_i) = \frac{\lambda_0}{2\pi n_i}\frac{d\Phi}{dx} \quad . \tag{4}$$

Equation (4) predicts that the reflected angle changes nonlinearly with the change of the incident angle. It is obviously different from the result of the classical reflection law. It should be addressed that there is a special incident angle, when the incident angle is larger than it the reflected light will disappear. This behavior is similar to the total internal reflection induced by refraction law. This critical angle can be defined as the critical angle for full transmission

$$\theta_{rc} = \arcsin(\pm 1 - \frac{\lambda_0}{2\pi n_i}\frac{d\Phi}{dx}) \quad . \tag{5}$$

From formula (5), it can be seen that the critical angle for full transmission always exists at the metasurface with an abrupt phase shift regardless of the both case that the light wave hits on the optical sparse material from the optical denser material (i.e. $n_t > n_i$), and vice versa (i.e. $n_t < n_i$). The critical angle for full transmission is determined by the refractive index $n_i$ of incident material, the wavelength of incident light $\lambda_0$ and the phase shift gradient $d\Phi/dx$ at the interface.

## 3. Derivation from the boundary conditions continuity

Although the general reflection and refraction laws have been derived, the interaction mechanism between light waves and the metasurface is still an open question. Whether the boundary continuity conditions are satisfied for the metasurface? Now, we make discussions.

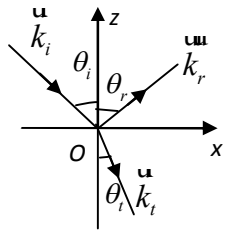

Fig. 2. Schematics used to derive the generalized laws of reflection and refraction. There exists the interface between the two media to introduce an abrupt phase shift $\Phi(\vec{r})$ to the light wave, which is a function of the position along the interface.

Considering the case, an incident plane wave is in the plane of $xoz$ (See Fig. 2), the $x$ axis is just at the interface of the two substances, the incident wave vector is $\vec{k}_i$, the incident angle is $\theta_i$, and the wave vectors of the reflected wave and refracted wave are $\vec{k}_r$ and $\vec{k}_t$ respectively, the reflected angle and the refracted angle are $\theta_r$ and $\theta_t$, respectively. Assuming that the tangential components of the electric field is continuous,

then we have

$$E_{ix} + E_{rx} = E_{tx} \tag{6}$$

where $E_{ix} = A_{ix}e^{-i[\omega_i t - k_i(l_i x + m_i y + n_i z)]}$, $E_{rx} = A_{rx}e^{-i[\omega_r t - k_r(l_r x + m_r y + n_r z) + \Phi(x)]}$, $E_{tx} = A_{tx}e^{-i[\omega_t t - k_t(l_t x + m_t y + n_t z) + \Phi(x)]}$ respond the x components of the electric fields of the incident, reflected and refracted waves, respectively. And $l_\alpha, m_\alpha, n_\alpha (\alpha = i, r, t)$ is three directional cosine of the wave vectors $\vec{k_i}$, $\vec{k_r}$ and $\vec{k_t}$, respectively.

At the interface ($z=0$), the incident wave vector can be denoted as $\vec{k_i} = k_i(\sin\theta_i, 0, \cos\theta_i)$. The boundary conditions should be satisfied at any time $t$ and at any point $(x,y,z)$, so we have $m_r = m_t = 0$. Then, the following relation could be obtained:

$$k_i l_i x = k_r l_r x + \Phi(x) = k_t l_t x + \Phi(x) \tag{7}$$

Substituted $l_r = \sin\theta_r$, $l_t = \sin\theta_t$ into Eq. (7), we have

$$\frac{\omega}{c} n_i \sin(\theta_i) x = \frac{\omega}{c} n_i \sin(\theta_r) x - \Phi(x) = \frac{\omega}{c} n_t \sin(\theta_t) x - \Phi(x) \quad . \tag{8}$$

At the interface, Eq. (8) can be expressed as

$$\frac{\omega}{c} n_i \sin(\theta_i) = \frac{\omega}{c} n_i \sin(\theta_r) - \frac{d\Phi(x)}{dx} = \frac{\omega}{c} n_t \sin(\theta_t) - \frac{d\Phi(x)}{dx} \quad . \tag{9}$$

Therefore, the general reflection law and the general refraction law identical to Eqs. (1) and (4) can be obtained.

$$\begin{cases} \sin\theta_r - \sin\theta_i = \dfrac{\lambda_0}{2\pi n_i} \dfrac{d\Phi(x)}{dx} & \text{reflection law} \\ n_t \sin\theta_t - n_i \sin\theta_i = \dfrac{\lambda_0}{2\pi} \dfrac{d\Phi(x)}{dx} & \text{refraction law} \end{cases} \tag{10}$$

It can be seen that the general reflection and refraction laws derived from Fermat's law and boundary conditions continuity for the interface with an abrupt phase shift are exactly the same. It means that the boundary conditions continuity is still valid in the interaction between light waves and metasurface.

4. Analysis and discussion

It should be pointed out that when the phase shift gradient satisfies $d\Phi/dx=0$, Eq. (10) turn into

$$\begin{cases} \theta_i = \theta_r & \text{reflection law} \\ \theta_t = \arcsin\left(\dfrac{n_i}{n_t} \sin\theta_i\right) & \text{refraction law} \end{cases} \tag{11}$$

It is the classical reflection and refraction laws. That is to say, the classical reflection and refraction laws are the special cases of the general reflection and refraction laws.

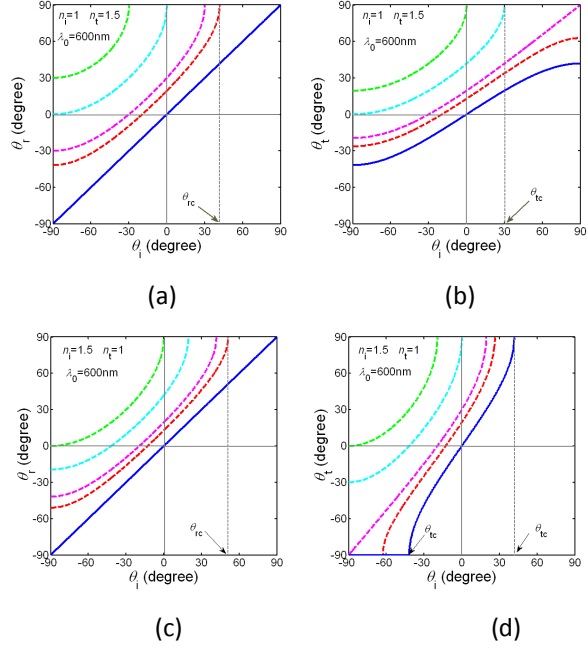

Fig. 3 The angles of reflection((a),and (c)) and refraction ((b),and (d)) governed by the traditional laws of reflection and refraction (blue solid line) and those given by the generalized laws of reflection and refraction (dotted line) change with the angle of incidence, when the light transmits from the optical sparse media into the optical denser medium ((a),and (b)) and the light incidents from the optical denser medium to the optical sparse medium ((c) and (d)), with the condition $d\Phi/dx = \pi/1.5\times10^7 \text{m}^{-1}$ (green dotted line), $d\Phi/dx = \pi/3\times10^7 \text{m}^{-1}$ (cyan dotted line), $d\Phi/dx = \pi/6\times10^7 \text{m}^{-1}$ (magenta dotted line), $d\Phi/dx = \pi/9\times10^7 \text{m}^{-1}$ (red dotted line).

By introducing the abrupt phase shift at the interface of the two materials, the reflected and refracted angles are related to the incidence angle, the refractive index of the two materials, the wavelength of incident light, and the phase shift gradient along the interface. Figure 3 demonstrates the curves of the reflected angles (Figs. 3(a) and 3(c)) with different incident angles, while Figs. 3(b) and 3(d) give the refracted angles changing with different incident angles. The blue solid line is the case determined by the classical reflection and refraction laws. Figs. 3(a) and 3(b) are the cases that the light propagates from the optical sparse material into the optical denser material, and Figs. 3(c) and 3(d) are the cases that the light transmits from the optical denser material to the optical sparse material. In our calculations, the wavelength in vacuum was chosen as $\lambda_0 = 600 \text{nm}$, the green, cyan, magenta and red dotted lines show the calculations with the phase shift gradient $d\Phi/dx = \pi/1.5\times10^7 \text{m}^{-1}$, $d\Phi/dx = \pi/3\times10^7 \text{m}^{-1}$, $d\Phi/dx = \pi/6\times10^7 \text{m}^{-1}$ and $d\Phi/dx = \pi/9\times10^7 \text{m}^{-1}$, respectively. The refraction indices $n_i = 1$, $n_t = 1.5$ were chosen in Figs. 3(a) and 3(b), and the value of the refraction indices $n_i = 1.5$, $n_t = 1$ were chosen in Figs. 3(c) and 3(d). From the Figs. 3(a) and 3(c), it can be seen that the relationship between the reflected angle and the incident angle given by the general reflection law is no longer linear. When the incident angle is larger than the critical angle $\theta_{rc}$, the reflected light disappears. The value of the critical angle for full transmission decreases gradually with the increase of the value of the phase change gradient $d\Phi/dx$ along the interface, reaches to zero, and then becomes negative. It means that there is a range of angles larger than 90º within the incidence plane in which there is no reflected light. This feature will be useful to design transmission components or ultra-thin film of antireflective coatings. From the Figs. 3(b) and 3(d), it is found that, when the light wave hits on the metasuface from a optical denser material to a optical sparse material, two different critical angle for total internal reflection exist at the condition $\frac{d\Phi}{dx} < \frac{\pi}{6} \times 10^7 \text{m}^{-1}$, and there is one critical angle for total internal reflection at the condition $\frac{d\Phi}{dx} > \frac{\pi}{6} \times 10^7 \text{m}^{-1}$. The results are consistent with the former analysis.

From Fig. 3, it can also be seen that the curves corresponding to the general reflection and refraction laws are all partially in the second quadrant. That is to say, the reflected light and the refracted light may locate at the same side of the normal of the interface with the incident light. Those are negative reflection and negative refraction, respectively. One can find some interesting phenomena at some particular incident angle. For example, ① the incident light may be opposite reflected at a certain nonzero incident angle for a appropriate interface which just like the case of normal incidence in classical reflection law; ② the refracted light may be in the same direction with the incident light at a nonzero incident angle which just likes no refraction. This kind of abnormal reflection and abnormal refraction can occur simultaneously if the wavelength of incident light, the refractive index of the two materials, and the phase shift gradient along the metasurface satisfy certain condition.

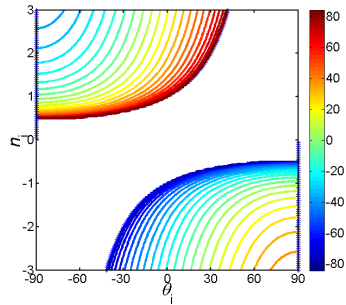

Fig. 4 The contour of the reflected angle with the change of changes the refractive index of the incident material and the incident angle at the condition of $\lambda_0 = 600 \text{nm}$ and $d\Phi / dx = \pi / 6 \times 10^7$.

Figure 4 gives the contour of the reflected angle for the different refractive indices of the incident material and the incident angle under the condition of $\lambda_0 = 600$ nm and $d\Phi / dx = \pi / 6 \times 10^7 \text{m}^{-1}$. It can be seen that the incident angle and the reflected angle are in nonlinear relation, and for a same incident angle the reflected angle decreases with the increase the refractive index of the incident material. In the contour, the curve in dark blue or in dark red represents case of the full transmission, in which the reflected light disappears. The critical angle of full transmission increases with increase of the refractive index of the incident material. This relation is caused by the assumption of $d\Phi/dx > 0$. It is consistent with the results of the Fig. 3.

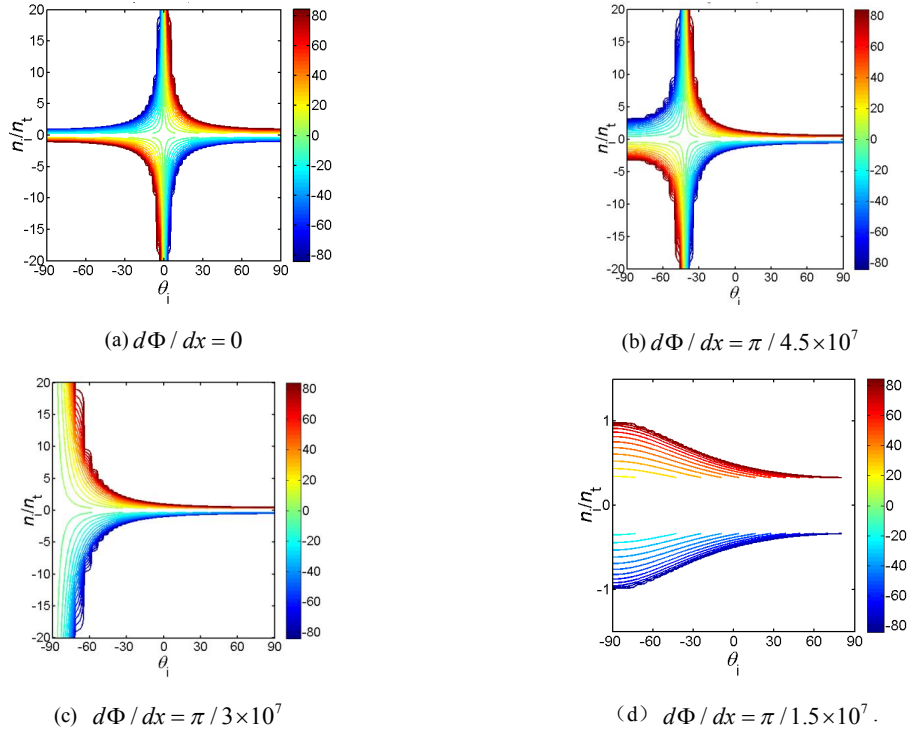

(a) $d\Phi/dx = 0$

(b) $d\Phi/dx = \pi/4.5\times10^7$

(c) $d\Phi/dx = \pi/3\times10^7$

(d) $d\Phi/dx = \pi/1.5\times10^7$.

Fig. 5 The contour of the refracted angle changes with the refractive index contrast of two media and the incident angle at the condition $\lambda_0 = 600$nm, $n_i = 1$,

Figure 5 gives the contour of the refracted angle changes with the incident angle and the refractive index contrast of two media for the case with the phase shift gradient along the metasurface $d\Phi/dx = 0$ (as shown in Fig. 5(a)), $d\Phi/dx = \pi/4.5\times10^7\,\text{m}^{-1}$ (as shown in Fig. 5(b)), $d\Phi/dx = \pi/3\times10^7\,\text{m}^{-1}$ (as shown in Fig. 5(c)), $d\Phi/dx = \pi/1.5\times10^7\,\text{m}^{-1}$ (as shown in Fig. 5(d)) with the condition $\lambda_0 = 600$ nm, $n_i = 1$, $n_t \in [-3,3]$. When the phase shift gradient along the metasurface is zero, the contours distribute symmetrically in the four quadrants. When the sign of the refraction index of the incident material is opposite to that of the refraction material, negative refraction will occur and the refraction angle will be spatially isotropic with regard to the incident angle, i.e, the optical path will be reversible in plane of incidence. However, when the phase gradient in the metasurface is not zero (as shown in Fig.5 (b)、Fig.5 (c) and Fig.5 (d)), the symmetry of the contours is broken. The optical path is no longer reversible in the plane of incidence. Also, negative refraction will not occur always for the case in which a light wave passes through two materials with opposite sign of the refractive indices (as shown in the third quadrant in Fig.5 (c)). While, negative refraction may occur for the cases in which the refractive indices of the two materials have same sign (as shown in the second quadrant in Fig.5 (c)). When the value of the phase gradient is $d\Phi/dx \geq \pi/3\times10^7\,\text{m}^{-1}$, the refracted angles are also positive for the case of $n_i n_t > 0$, and the refracted angles are also negative value for the cases of $n_i n_t < 0$. The condition of these cases can be denoted as $\text{sgn}(n_i n_t) = \text{sgn}(\theta_t)$ regardless of the sign of the incident angle. The dividing line of positive refraction and negative refraction is the normal of the interface of the two materials regardless of their optical properties. It should be noted that, there is a critical value $R_c$ of the ratio $n_i / n_t$ for a particular phase gradient with the condition $d\Phi / dx > \pi / 3 \times 10^7 \, \text{m}^{-1}$, when $n_i / n_t > R_c$, the refracted wave disappears. The value of $R_c$ just corresponds to the critical angle of total internal reflection and is determined by the formula (2).

In the above discussions, it is assumed that the abrupt phase shift is a continuous function of the position along the metasurface, and the phase shift gradient $d\Phi / dx$ is a constant. Therefore, the transfer of the energy of the incident wave cross the metasurface is determined by the general reflection and refraction laws regardless of the polarization state of incident light. This phenomenon of abnormal refraction caused by the discontinuity of the phase shift along the interface is quite different from the relation established in metamaterials with the negative permittivity, negative permeability and the anisotropy of the permittivity tensor[15].

## 5. Design of the metasurface

The general reflection and refraction laws can be used to design elements to control the wave front by introducing the abrupt phase shift in the optical path. V-shaped antenna array [15,16] had been designed to achieve the abrupt phase shift along the interface based on the surface plasma resonance, and verified experimentally by reorganizing the plasma antenna unit. However, the V-shaped antenna can only generate the abrupt phase shift for the cross-polarized scattered light, but still, the original polarization scattering light can not be controlled.

If a metasurface generating the abrupt phase shift for arbitrary polarization direction is designed, the reflected light and refracted light will be controlled to achieve arbitrary manipulation. In fact, a phase mask can be used as the metasurface. Assuming that the interface is a one-dimensional phase grating which can change abruptly the phase of the light wave. The thickness of the grating is about tens of nanometers, its transmission function is $t(x) = t_0 e^{i\phi(x)}$, and the phase shift is $\phi(x) = 2\pi x / T$ along the direction of the grating vector, $T$ is the period of the grating. So, we have $d\phi / dx = 2\pi / T$. The thickness of such a structure is much smaller than the wavelength of light, thus, it can be treated as a kind of metasurface. Pasting the metasurface on a glass surface, we have $n_i = 1$, $n_t = 1.5$. When a light with the wavelength of $\lambda_0$ passes through the metasurface at a incident angle of $\theta_i$, the refracted angle can be written as $\sin(\theta_t) = \sin(\theta_i) + \dfrac{2\lambda_0}{3T}$. Usually, the optical phase can be changed by changing the geometry path of light and/or the refractive index of the material.

The numerical simulations were done using the Comsol software to study the propagation behavior of light crossing the ultra-thin interface. In our simulations, the refractive index gradually increased along the x-axis direction (See Fig. 1) with $dn/dx = 1/60 \text{ nm}^{-1}$, while the thickness was kept unchanged. The computational domain was $1.8\,\mu\text{m} \times 2\,\mu\text{m}$, the thickness of the metasurface was $30\,\text{nm}$, and $n_i = 1$, $n_t = 1.5$ were chosen. In order to reflect the variation of the refractive index along the metasurface, the cross section of the whole interface between the air and the glass was divided into sub-areas with the size of $30\,\text{nm} \times 30\,\text{nm}$, the difference of the refractive indices between the two adjacent sub-areas was $\Delta n = 0.5$, the frequency of the light was chosen as $f = 10 \times 10^{14}\,\text{Hz}$ ($\lambda_0 = 0.3\,\mu\text{m}$), and the refractive index at right side was larger than that at left side. The metasurface can change the phase of the light wave with $d\phi/dx = \pi/\lambda_0$. Figure 6 shows the simulating results at different incident angles. It can be seen that different refracted angle corresponds to different incident angle, and negative refraction may occur (See Fig. 6(c)). These results satisfy the relation $1.5\sin(\theta_t) = \sin(\theta_i) + \dfrac{1}{2}$ and it verified the previous results. It should be pointed out that the method for impedance matching was employed to minimize the affection of reflection so that to well demonstrate the simulating results. Similar results were obtained for other wavelengths in our simulations.

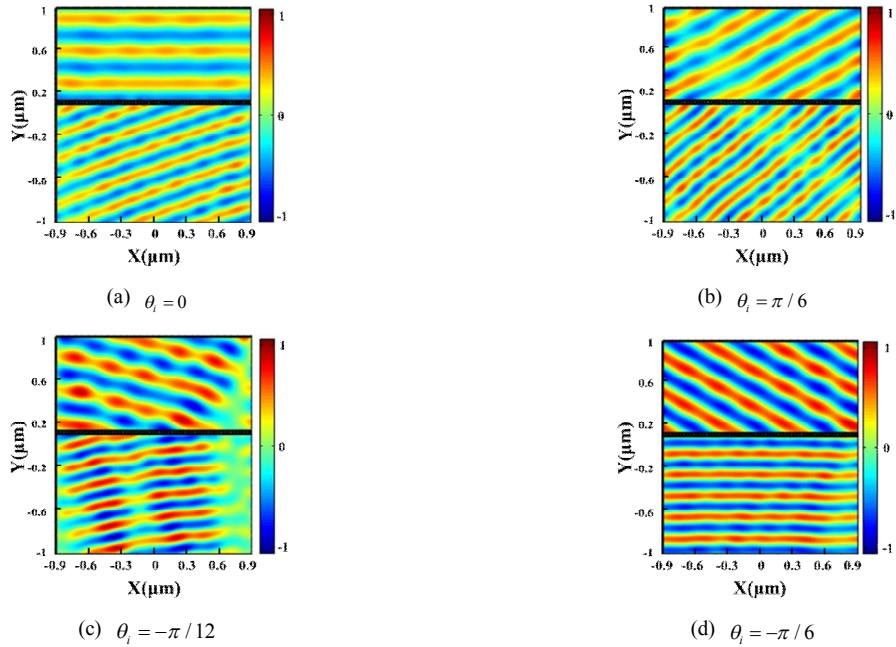

(a) $\theta_i = 0$      (b) $\theta_i = \pi/6$

(c) $\theta_i = -\pi/12$      (d) $\theta_i = -\pi/6$

Fig. 6 The simulation results when the light waves with frequency $f = 10 \times 10^{14}\,\text{Hz}$ at different angles

In the above simulations, it was assumed that the refractive index changes linearly along the metasurface.

Several techniques can be used to achieve refractive index change, such as electro-optical effect, accusto-optical effect, in designing a variety of ultra-thin optical devices.

**Acknowledgements**

The authors thank the National Natural Science Foundation of China (grant Nos. 11074024 and 11104016) and the Specialized Research Fund for the Doctoral Program of Higher Education (SRFDP) Education ministry of China for the financial support. We also would like to thank Dr. Jun Zheng for her help with the simulations.